\documentclass[aps,prd,reprint,groupedaddress,showkeys]{revtex4-2}
\usepackage[utf8]{inputenc}
\usepackage[english]{babel}
\usepackage[
  bookmarks=false,
  pdfpagelabels=false,
]{hyperref}
\usepackage{color}
\usepackage{amsmath}
\usepackage{amsfonts}
\usepackage{amssymb}
\usepackage{graphicx}
\usepackage[left=2cm,right=2cm,top=2cm,bottom=2cm]{geometry}
\newcommand{\dx}{\mathrm{d}}
\newcommand{\eps}{\varepsilon}
\newcommand{\dilog}{\mathrm{Li}_2}
\newcommand{\trilog}{\mathrm{Li}_3}
\newcommand{\doublenielsentrilog}{\mathrm{S}_{1,1,1}}
\begin{document}

\preprint{}

\title{Asymptotic behavior of angular integrals in the massless limit}

\author{Fabian Wunder}%\,\orcidlink{0009-0007-4136-7844}}
\email[]{fabian.wunder@uni-tuebingen.de}
%\homepage[]{Your web page}
%\thanks{}
%\altaffiliation{}
\affiliation{Institute for Theoretical Physics, University of T\"ubingen\\
Auf der Morgenstelle 14, 72076 T\"ubingen, Germany}

\date{\today}

\begin{abstract}
We investigate the small-mass asymptotics of a class of massive $d$ dimensional angular integrals.
These integrals arise in a wide range of perturbative quantum field theory calculations.
We derive expressions characterizing their behavior in the vicinity of the massless limit for all cases with up to two denominators.
The results established in this work are applicable to phase-space calculations where an integration over virtuality including the massless limit is required.
\end{abstract}

% insert suggested keywords - APS authors don't need to do this

\maketitle

\section{Introduction}
\label{sec: Introduction}
Angular integrals \cite{Schellekens:1981,vanNeerven:1985,Beenakker:1988,Somogyi:2011,Lyubovitskij:2021} are ubiquitous to phase-space calculations in perturbative quantum field theory \cite{Bolzoni:2010,Anastasiou:2013, Lillard:2016,Kotlarski:2016,Lionetti:2018, Specchia:2018,Bahjat-Abbas:2018, Baranowski:2020,Blumlein:2020,Isidori:2020,Alioli:2022,Assi:2023,Catani:2023,Pal:2023,Devoto:2024}.
Examples from QCD include theoretical predictions for the Drell-Yan process (DY) \cite{Matsuura:1989,Matsuura:1990,Hamberg:1991,Mirkes:1992,Bahjat-Abbas:2018}, deep-inelastic scattering (DIS) \cite{Duke:1982,Hekhorn:2019}, semi-inclusive deep-inelastic scattering (SIDIS) \cite{Anderle:2016,Wang:2019}, prompt-photon production \cite{Gordon:1993}, hadron-hadron scattering \cite{Ellis:1980}, heavy quark production \cite{Beenakker:1988}, and single-spin asymmetries \cite{Schlegel:2012, Ringer:2015}.

When massless particles are present, the angular integration contains collinear singularities.
To regularize these divergencies, the calculations are performed in $d=4-2\eps$ dimensions \cite{tHooft:1972,Bollini:1972}.

Following the notation from references \cite{Somogyi:2011,Lyubovitskij:2021} we define the angular integral with two denominators as
\begin{equation}
I^{(m)}_{j_1,j_2}(v_{12},v_{11},v_{22};\eps)\equiv\int\dx\Omega_k\,\frac{1}{(v_1\cdot k)^{j_1}\,(v_2\cdot k)^{j_2}}
\label{eq: definition of angular integral}
\end{equation}
with normalized $d$-vectors
\begin{align*}
k&=(1,\dots,\sin\theta_1\cos\theta_2,\cos\theta_1)\,,
\\
v_1&=(1,\bold{0}_{d-2},\beta_1)\,,
\\
v_2&=(1,\bold{0}_{d-3},\beta_2\sin\chi,\beta_2\cos\chi)\,,
\end{align*}
kinematic invariants $v_{ij}=v_i\cdot v_j$,
and integration measure ${\dx\Omega_k = \dx\theta_1 \sin^{1-2\eps} \theta_1\dx\theta_2 \sin^{-2\eps} \theta_2}$.
The denominator powers $j_1$, $j_2$ are assumed to be integers in the following.
The superscript $m=0,1,2$ characterizes the number of non-zero \textit{masses} $v_{11}$, $v_{22}$.
For convenience, zero indices and masses will be dropped from the notation, i.e. we will write for example $I^{(1)}_{j_1}(v_{11};\eps)$ instead of $I^{(1)}_{j_1,0}(v_{12},v_{11},v_{22};\eps)$ and $I^{(0)}_{j_1,j_2}(v_{12};\eps)$ instead of $I^{(0)}_{j_1,j_2}(v_{12},0,0;\eps)$.
By partial fraction decomposition a wide range of phase-space integrals can be cast into the form of Eq{.}\,\eqref{eq: definition of angular integral} \cite{Duke:1982,Lyubovitskij:2021}.

In this manuscript we investigate the asymptotic behavior of integrals of the form $I^{(m)}_{j_1,j_2}(v_{12},v_{11},v_{22};\eps)$ in the limit of one or both masses going to zero.
In principle, the expansion of all two-denominator angular integrals with integer powers $j_1,j_2$ is known to all orders in the dimensional regularization parameter $\eps$ \cite{Beenakker:1988,Somogyi:2011,Lyubovitskij:2021}.
However, these expansions are not always sufficient.

As an illustration of the potential issue occurring in the massless limit, let us look at the double-massive angular integral with $j_1=j_2=1$.
It has the well known $\eps$-expansion \cite{Schellekens:1981,Beenakker:1988,Somogyi:2011,Lyubovitskij:2021}
\begin{equation}
\!I^{(2)}_{1,1}(v_{12},v_{11},v_{22};\eps) = 
\frac{\pi}{\sqrt{X}}\log\!\left(\frac{v_{12}+\sqrt{X}}{v_{12}-\sqrt{X}}\right)+\mathcal{O}(\eps)\,,
\label{eq: double massive angular integral plain expansion}
\end{equation}
with $X=v_{12}^2-v_{11} v_{22}$.
%The expansion is known to all orders in $\eps$ in terms of double Nielsen polylogarithms \cite{Lyubovitskij:2021}.

One readily sees that the massless limit $v_{11}\rightarrow 0$ is ill-defined at the level of the $\eps$-expansion, since $v_{12}-\sqrt{X}$ approaches zero.
This is a problem if we were to consider an integral of the form
\begin{align}
\int_0^{v_{11}^\text{max}}\dx v_{11}\,v_{11}^{-1-\eps}\,I^{(2)}_{1,1}(v_{12},v_{11},v_{22};\eps)\,.
\label{eq: singular v11 integration}
\end{align}
Here, we would like to replace $I^{(2)}_{1,1}(v_{12},v_{11},v_{22};\eps)$ by its $\eps$-expansion under the integral and employ the distributional identity \cite{Altarelli:1979,Campbell:2017,Haug:2022}
\begin{align}
\!\!v_{11}^{-1-n\eps}=-\frac{1}{n\eps}\,\delta(v_{11})+\sum_{n=0}^\infty \frac{(-n\eps)^n}{n!} \left[\frac{\log^n v_{11}}{v_{11}}\right]_+
\label{eq:Plus distribution expansion}
\end{align}
on $v_{11}^{-1-\eps}$.
However, we cannot use the form of Eq{.}\,\eqref{eq: double massive angular integral plain expansion} due to its divergence in the $v_{11}\rightarrow 0$ limit.
Instead, to properly perform the integration one has to extract the asymptotic behavior of $I^{(2)}_{1,1}(v_{12},v_{11},v_{22};\eps)$ near $v_{11}=0$ beforehand, resulting in additional powers of $v_{11}^{-\eps}$ entering Eq{.}\,\eqref{eq:Plus distribution expansion}.

The aim of this work is to provide $\eps$-expansions for all massive angular integrals with up to two propagators, where the asymptotic behavior in the massless limit is manifest and which are hence suitable for usage within integrals of the form \eqref{eq: singular v11 integration}.

Using recursion relations derived from integration-by-parts (IBP) identities, the powers $j_1$ and $j_2$ can always be reduced to the cases $j_{1,2}=0$ or $1$ \cite{Lyubovitskij:2021}.
Hence it suffices to consider the master integrals $I^{(1)}_1(v_{11};\eps)$, $I^{(1)}_{1,1}(v_{12},v_{11};\eps)$, and $I^{(2)}_{1,1}(v_{12},v_{11},v_{22};\eps)$.

The remainder of this manuscript is organized as follows.
In section \ref{sec: Asymptotic behavior in the massless limit}{.} we recall the two-point splitting lemma which we subsequently use to establish the asymptotic behavior of the master integrals in the massless limits $v_{11},v_{22}\rightarrow 0$.
Section \ref{sec: Conclusion}{.} concludes the paper.

\section{Asymptotic behavior in the massless limit}
\label{sec: Asymptotic behavior in the massless limit}
The main tool for the extraction of the asymptotic behavior will be the \textit{two-point splitting lemma} \citep{Lyubovitskij:2021}.
Using the notation
\begin{align}
\Delta_k(v_i,v_j)\equiv\frac{1}{v_i\cdot k\,\,v_j\cdot k},
\end{align}
it states that for any two vectors $v_1$ and $v_2$, we can choose any scalar $\lambda$ and construct the linear combination
${
v_3=(1-\lambda)v_1+\lambda v_2
}$
to obtain the identity
\begin{align}
\Delta_k(v_1,v_2)=\lambda\,\Delta_k(v_1,v_3)+(1-\lambda)\,\Delta_k(v_2,v_3)\,.
\label{eq: Two point splitting lemma}
\end{align}

This allows us to express a given angular integral in terms of other angular integrals where a new auxiliary vector $v_3$ has been inserted.
By choosing appropriate values for $\lambda$, the vector $v_3$ can be given desirable properties, most importantly being massless.
This idea has been fruitfully employed in reference \citep{Lyubovitskij:2021} for the calculation of the all-order $\eps$-expansion of the double-massive integral.
\subsection{Asymptotic form of the massive one-denominator integral}
\label{sec:Asymptotic form of the massive one denominator integral}
We start with the investigation of the massive one-denominator master integral
\begin{equation}
I^{(1)}_1(v_{11};\eps)=\int\dx\Omega_k\,\frac{1}{v_1\cdot k}\,.
\label{eq: massive one denominator integral}
\end{equation}
Its $\eps$-expansion is \cite{Beenakker:1988,Somogyi:2011,Lyubovitskij:2021}
\begin{align}
I^{(1)}_1(v_{11};\eps)=\frac{\pi}{\sqrt{1-v_{11}}}\,\log\!\left(\frac{1+\sqrt{1-v_{11}}}{1-\sqrt{1-v_{11}}}\right)+\mathcal{O}(\eps)\,,
\label{eq: massive one denominator integral plain expansion}
\end{align}
which is singular in the $v_{11}\rightarrow 0$ limit.

\begin{figure}
\includegraphics[width=0.87\columnwidth]{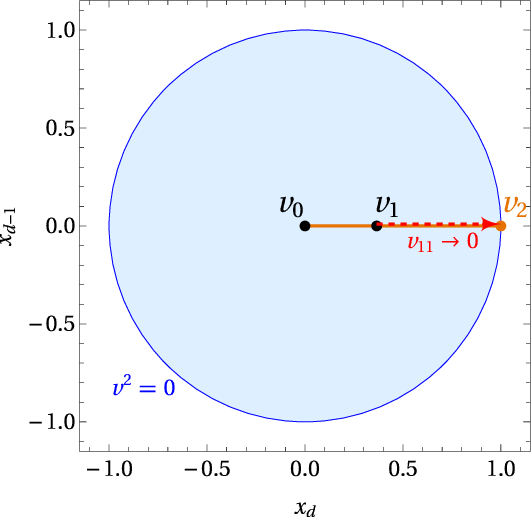}
\caption{Sketch illustrating the splitting of the massive one-denominator integral in Eq{.}\,\eqref{eq: Two point splitting lemma massive one denominator integral}. The figure shows the slice $x_0=1$ of Minkowski space, the blue circle indicates the intersection with the light-cone where $v^2=0$.\label{fig:Massive one denominator splitting}}
\end{figure}

To extract the massless limit from Eq{.}\,\eqref{eq: massive one denominator integral} explicitly, we define the auxiliary ``zero" vector $v_0=(1,\bold{0}_{d-1})$ with ${1/(v_0\cdot k)=1}$ and set ${v_2=(1-\lambda) v_0+\lambda v_{1}}$.
Demanding $v_2$ to be massless, i.e. $v_{22}=0$, we find $\lambda=1/\sqrt{1-v_{11}}$.
We observe that $v_1$ indeed approaches $v_2$ in the limit $v_{11}\rightarrow 0$.
A graphical illustration of this construction is given in Fig{.}\,\ref{fig:Massive one denominator splitting}. 

The two-point splitting lemma \eqref{eq: Two point splitting lemma} provides us with the identity
\begin{align}
\Delta_k(v_0,v_1)=\lambda\,\Delta_k(v_0,v_2)+(1-\lambda)\,\Delta_k(v_1,v_2)\,.
\label{eq: Two point splitting lemma massive one denominator integral}
\end{align}
Integrating Eq{.}\,\eqref{eq: Two point splitting lemma massive one denominator integral} and substituting the value for $\lambda$ we get
\begin{equation}
I_{1}^{(1)}(v_{11};\eps)=\frac{I_{1}^{(0)}(\eps)}{\sqrt{1-v_{11}}}\!-\!\frac{1-\sqrt{1-v_{11}}}{\sqrt{1-v_{11}}}I_{1,1}^{(1)}(v_{12},v_{11};\eps).
\label{eq: Splitting of massive one denominator integral}
\end{equation}

Hence, we have transformed the massive one-denominator integral into the sum of a massless one-denominator integral and a single-massive two-denominator integral, where the coefficient of the latter vanishes in the massless limit.
It is $ I_{1}^{(0)}(\eps)=-\pi/\eps$ and the $\eps$-expansion of the single-massive two-denominator integral is \citep{Beenakker:1988,Somogyi:2011,Lyubovitskij:2021}
\begin{align}
&I_{1,1}^{(1)}(v_{12},v_{11};\eps)=\frac{\pi}{v_{12}}\left(\frac{v_{11}}{v_{12}^2}\right)^\eps
\nonumber\\
&\quad\times\left[-\frac{1}{\eps}-2\eps\,\left(\dilog(\omega_{12}^+)+\dilog(\omega_{12}^-)\right)+\mathcal{O}(\eps^2)\right],
\label{eq: single massive two denominator integral plain expansion}
\end{align}
with $\omega^\pm_{12}=1-v_{12}/(1\pm\sqrt{1-v_{11}})$.
For the one-denominator kinematics we have ${v_{12}=1-\sqrt{1-v_{11}}}$, ${\omega^+_{12}=2\sqrt{1-v_{11}}/(1+\sqrt{1-v_{11}})}$, and ${\omega^-_{12}=0}$.

Plugging the $\eps$-expansions into Eq{.}\,\eqref{eq: Splitting of massive one denominator integral}, we receive
\begin{align}
&I_{1}^{(1)}(v_{11};\eps)=-\frac{\pi}{\sqrt{1-v_{11}}}
\left\lbrace\frac{1}{\eps}+v_{11}^{-\eps}(1+\sqrt{1-v_{11}})^{2\eps}
\right.
\nonumber\\
&\quad\times\left.\left[-\frac{1}{\eps}-2\eps\,\dilog\left(\frac{2\sqrt{1-v_{11}}}{1+\sqrt{1-v_{11}}}\right)+\mathcal{O}(\eps^2)\right]\right\rbrace.
\label{eq: Asymptotic form of one denominator integral}
\end{align}
In this form the asymptotic behavior for $v_{11}\rightarrow 0$ is explicit.
We observe that $I^{(1)}_{1}(v_{11};\eps)$ has a part constant in the massless limit and a part proportional to $v_{11}^{-\eps}$.
It is the latter that causes the logarithmic divergence in Eq{.}\,\eqref{eq: massive one denominator integral plain expansion}.
Note that both parts have a $1/\eps$ pole which cancels between the two for finite $v_{11}$.
\subsection{Asymptotic form of the single-massive two-denominator integral}
\label{sec:Asymptotic form of the single massive two denominator integral}
The second master integral we look at, is the single-massive two-denominator integral
\begin{equation}
I^{(1)}_{1,1}(v_{12},v_{11};\eps)=\int\dx\Omega_k\,\frac{1}{v_1\cdot k\,\,v_2\cdot k}\,.
\label{eq: single massive two denominator integral}
\end{equation}
We have already encountered its $\eps$-expansion in Eq{.}\,\eqref{eq: single massive two denominator integral plain expansion}.
We observe that the expansion is singular in the limit $v_{11}\rightarrow 0$ because the variable $\omega_{12}^{-}$ diverges.

\begin{figure}
\includegraphics[width=0.87\columnwidth]{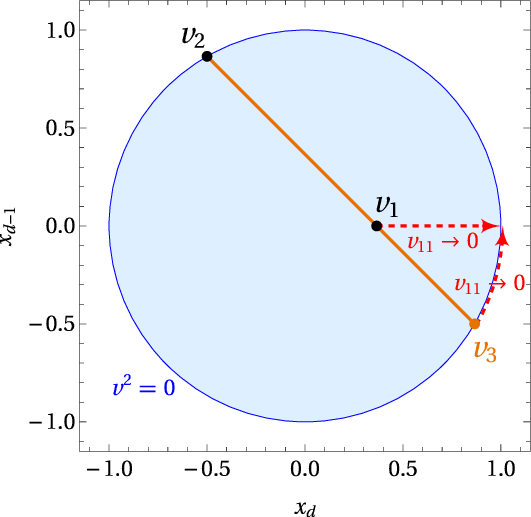}
\caption{Sketch illustrating the splitting of the single-massive two-denominator integral in Eq{.}\,\eqref{eq: Splitting of single massive two denominator integral}. The figure shows the slice $x_0=1$ of Minkowski space, the blue circle indicates the intersection with the light-cone where $v^2=0$.\label{fig:Single massive two denominator splitting}}
\end{figure}

To extract the asymptotic behavior of the single-massive integral for small masses, we want to separate off its massless limit.
To this end we define the auxiliary vector ${v_3=(1-\lambda)\,v_1+\lambda\,v_2}$.
For $v_{3}$ to be massless, i.e. $v_{33}=0$, we set ${\lambda=v_{11}/(v_{11}-2 v_{12})}$.
A graphical illustration of the splitting construction is given in Fig{.}\,\ref{fig:Single massive two denominator splitting}. 

Upon integration of the associated two-point splitting identity, which is of the form of Eq{.}\,\eqref{eq: Two point splitting lemma}, we receive
\begin{align}
I_{1,1}^{(1)}(v_{12},v_{11};\eps)=
\frac{2v_{12} I_{1,1}^{(0)}(v_{23};\eps)}{2v_{12}-v_{11}}+
\frac{v_{11} I_{1,1}^{(1)}(v_{13},v_{11};\eps)}{v_{11}-2v_{12}}
\,,
\label{eq: Splitting of single massive two denominator integral}
\end{align}
with the scalar products ${v_{13}=v_{11}v_{12}/(2v_{12}-v_{11})}$ and ${v_{23}=2v_{12}^2/(2v_{12}-v_{11})}$.

Hence, we have transformed the single-massive two-denominator integral into the sum of a massless two-denominator integral and a single-massive two-denominator angular integral, where the coefficient of the latter vanishes in the massless limit.

The $\eps$-expansion of the massless two-denominator integral reads \citep{Beenakker:1988,Somogyi:2011,Lyubovitskij:2021}
\begin{align}
I_{1,1}^{(0)}(v_{12};\eps)&=\pi\left(\frac{v_{12}}{2}\right)^{-1-\eps}
\nonumber\\
&\times\left[-\frac{1}{\eps}-\eps\,\dilog\left(1-\frac{v_{12}}{2}\right)+\mathcal{O}(\eps^2)\right],
\label{eq: massless two denominator integral expansion}
\end{align}
for the expansion of $I_{1,1}^{(1)}(v_{13},v_{11};\eps)$ we can again use Eq{.}\,\eqref{eq: single massive two denominator integral plain expansion}.
Plugging these into \eqref{eq: Splitting of single massive two denominator integral}, we receive
\begin{align}
&I_{1,1}^{(1)}(v_{12},v_{11};\eps)=-\frac{2\pi}{v_{12}}
\nonumber\\
&\!\!\times\left\lbrace
v_{12}^{-\eps}(2\nu)^{\eps}
\left[
\frac{1}{\eps}
+\eps\left(
\dilog\!\left(1-\frac{v_{12}}{2\nu}\right)
\right)+\mathcal{O}(\eps^2)
\right]
\right.
\label{eq: Asymptotic form of single massive integral}
\\
&\!\!\left.
-v_{11}^{-\eps}(2\nu)^{2\eps}\left[
\frac{1}{2\eps}
+\eps\left(\dilog(\omega_{13}^+)+\dilog(\omega_{13}^-)
\right)+\mathcal{O}(\eps^2)
\right]
\right\rbrace,
\nonumber
\end{align}
%widetext layout of equation
%\begin{widetext}
%\begin{align}
%I_{1,1}^{(1)}(v_{12},v_{11};\eps)=&-\frac{2\pi}{v_{12}}
%\left\lbrace
%v_{12}^{-\eps}\left[
%\frac{1}{\eps}+\log\!\left(2-\frac{v_{11}}{v_{12}}\right)+\eps\left(
%\dilog\!\left(1-\frac{v_{12}^2}{2v_{12}-v_{11}}\right)+\frac{1}{2}\log^2\!\left(2-\frac{v_{11}}{v_{12}}\right)
%\right)+\mathcal{O}(\eps^2)
%\right]
%\right.
%\nonumber\\
%&\phantom{\left.\frac{2\pi}{v_{12}}\right\rbrace}\,\left.
%-v_{11}^{-\eps}\left[
%\frac{1}{2\eps}+\log\!\left(2-\frac{v_{11}}{v_{12}}\right)+\eps\left(\dilog(\omega_{13}^+)+\dilog(\omega_{13}^-)+\log^2\!\left(2-\frac{v_{11}}{v_{12}}\right)
%\right)+\mathcal{O}(\eps^2)
%\right]
%\right\rbrace,
%\end{align}
%\end{widetext}
with ${\omega_{13}^\pm=(v_{12}(1\pm\sqrt{1-v_{11}})-v_{11})/(2v_{12}-v_{11})}$ and the abbreviation ${\nu=1-v_{11}/(2v_{12})}$.
In the massless limit $\omega_{13}^\pm$ approaches $v_{12}$ respectively $0$, and $\nu$ goes to $1$.

Again we have found a form of the $\eps$-expansion, where the asymptotic behavior for $v_{11}\rightarrow 0$ is explicit.
As for the one-denominator integral, we have a finite part and a part proportional to $v_{11}^{-\eps}$.
Note that Eq{.}\,\eqref{eq: Asymptotic form of single massive integral} trivially reduces to Eq{.}\,\eqref{eq: massless two denominator integral expansion} for $v_{11}=0$, something that could not be easily seen from Eq{.}\,\eqref{eq: single massive two denominator integral plain expansion}.

\subsection{Asymptotic form of the double-massive two-denominator integral}
\label{sec:Asymptotic form of the double massive two denominator integral}
Finally, we consider the double-massive two-denominator master integral,
\begin{equation}
I^{(2)}_{1,1}(v_{12},v_{11},v_{22};\eps)=\int\dx\Omega_k\,\frac{1}{v_1\cdot k\,\,v_2\cdot k}\,.
\label{eq: double massive two denominator integral}
\end{equation}
We have already discussed the divergent behavior of its $\eps$-expansion in the introduction, see Eq{.}\,\eqref{eq: double massive angular integral plain expansion}.

Using two-point splitting, the double-massive integral can be expressed as a sum of single-massive integrals \citep{Lyubovitskij:2021}.
For the double-massive integral, we have to consider the cases of one or both masses approaching zero.
To treat both limits together we will employ a splitting that treats $v_1$ and $v_2$ symmetrically and directly extracts the double massless limit.

\begin{figure}
\includegraphics[width=0.87\columnwidth]{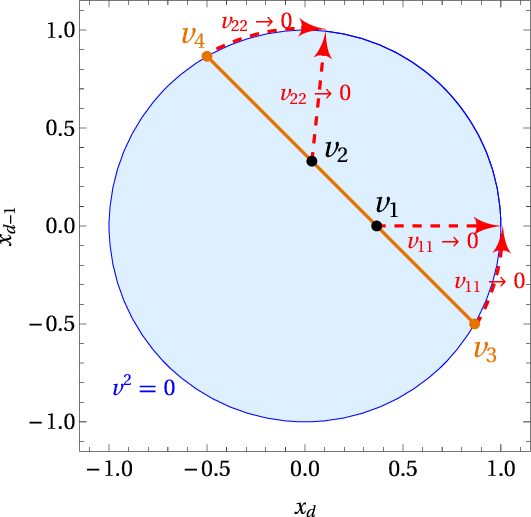}
\caption{Sketch illustrating the splitting of the double-massive two-denominator integral in Eq{.}\,\eqref{eq: Two point splitting lemma double massive two denominator integral}. The figure shows the slice $x_0=1$ of Minkowski space, the blue circle indicates the intersection with the light-cone where $v^2=0$.\label{fig:Double massive two denominator splitting}}
\end{figure}

We define two auxiliary vectors ${v_3=(1-\lambda)v_1+\lambda v_2}$ and ${v_4=\mu\,v_1+(1-\mu)\,v_2}$.
Employing the two-point splitting lemma \,\eqref{eq: Two point splitting lemma} first on $\Delta_k(v_1, v_2)$ inserting $v_3$ and subsequently on $\Delta_k(v_2, v_3)$ inserting $v_4$, we obtain the splitting
\begin{align}
\Delta_k(v_1,v_2)=&\lambda\,\Delta_k(v_1, v_3)+\mu\,\Delta_k(v_2, v_4)
\nonumber\\
&+(1-\lambda-\mu)\,\Delta(v_3,v_4)\,.
\label{eq: Two point splitting lemma double massive two denominator integral}
\end{align}
To make $v_3$ and $v_4$ massless as well as coinciding with $v_1$ respectively $v_2$ in the respective massless limit, we choose
\begin{align}
\lambda=\frac{v_{12}-v_{11}-\sqrt{X}}{2v_{12}-v_{11}-v_{22}}\,,\quad
\mu=\frac{v_{12}-v_{22}-\sqrt{X}}{2v_{12}-v_{11}-v_{22}}\,,
\end{align}
with $X=v_{12}^2-v_{11}v_{22}$.
The scalar products of the auxiliary vectors are ${v_{33}=v_{44}=0}$, ${v_{13}=-\lambda\sqrt{X}}$, ${v_{24}=-\mu\sqrt{X}}$, and ${v_{34}=2X/(2v_{12}-v_{11}-v_{22})}$.
Note that both $v_{13}$ and $v_{24}$ vanish in the respective massless limits.
A graphical illustration of the splitting is given in Fig{.}\,\ref{fig:Double massive two denominator splitting}. 

Upon integration of Eq{.}\,\eqref{eq: Two point splitting lemma double massive two denominator integral}, we receive
\begin{align}
&I_{1,1}^{(2)}(v_{12},v_{11},v_{22};\eps)=\frac{\pi}{\sqrt{X}}\left[v_{34}\,I_{1,1}^{(0)}(v_{34};\eps)
\right.
\nonumber\\
&\left.\qquad-v_{13}\,I_{1,1}^{(1)}(v_{13},v_{11};\eps)
-v_{24}\,I_{1,1}^{(1)}(v_{24},v_{22};\eps)\right].
\label{eq: Splitting of double massive two denominator integral}
\end{align}
This identity splits the double-massive integral into two single-massive integrals and a massless integral.
For these we can use the $\eps$-expansions from eqs.~\eqref{eq: single massive two denominator integral plain expansion} and \eqref{eq: massless two denominator integral expansion},
resulting in the representation of the double-massive integral
\begin{align}
\label{eq: Asymptotic form of double massive integral}
&I_{1,1}^{(2)}(v_{12},v_{11},v_{22};\eps)=\frac{\pi}{\sqrt{X}}
\left\lbrace
2\left(\frac{v_{34}}{2}\right)^{-\eps}
\right.
\nonumber
\\
&\quad\times\left.
\left[-\frac{1}{\eps}-\eps\,\dilog\left(1-\frac{v_{34}}{2}\right)+\eps^2 f_0(v_{34})+\mathcal{O}(\eps^3)\right]
\right.
\nonumber\\
&\quad\left.
-v_{11}^{-\eps}\left(\frac{v_{11}}{v_{13}}\right)^{2\eps}
\left[-\frac{1}{\eps}-2\eps\,\left(\dilog(\omega_{13}^+)+\dilog(\omega_{13}^-)\right)
\right.\right.
\nonumber\\
&\qquad\left.\left.+\eps^2 f_1(\omega_{13}^+,\omega_{13}^-)+\mathcal{O}(\eps^3)\vphantom{\frac{1}{\eps}}\right]
\right.
\\
&\quad\left.
-v_{22}^{-\eps}\left(\frac{v_{22}}{v_{24}}\right)^{2\eps}
\left[-\frac{1}{\eps}-2\eps\,\left(\dilog(\omega_{24}^+)+\dilog(\omega_{24}^-)\right)
\right.\right.
\nonumber\\
&\qquad\left.\left.+\eps^2 f_1(\omega_{24}^+,\omega_{24}^-)+\mathcal{O}(\eps^3)\vphantom{\frac{1}{\eps}}\right]
\right\rbrace,
\nonumber
\end{align}
where ${\omega_{ij}^\pm=1-v_{ij}/(1\pm\sqrt{1-v_{ii}})}$.
The kinematic variables $X$, $v_{13}$, $v_{24}$, and $v_{34}$ are defined in the text above; they all depend on $v_{12}$, $v_{11}$, and $v_{22}$.
Importantly $v_{11}/v_{13}\rightarrow 2$ for $v_{11}\rightarrow 0$ and analogously $v_{22}/v_{23}\rightarrow 2$ for $v_{22}\rightarrow 0$.

The asymptotics of the double-massive integral is manifest in Eq{.}\,\eqref{eq: Asymptotic form of double massive integral}, we have a part constant in both massless limits, a part proportional to $v_{11}^{-\eps}$, and a part proportional to $v_{22}^{-\eps}$.
Note that the $1/\eps$ poles cancel between the parts if we expand in $\eps$ for finite $v_{11}$ and $v_{22}$.
For $v_{22}=0$ we immediately recover Eq{.}\,\eqref{eq: Asymptotic form of single massive integral}.

The full expressions for the functions $f_{0,1}$ parametrizing the order $\eps^2$ parts of the massless respectively single-massive two-denominator integral can be found in the appendix\ref{app: order eps^2 coefficients}.
The order $\eps^2$ is included here, since applying the expansion \eqref{eq:Plus distribution expansion} for both $v_{11}$ and $v_{22}$ may result in a $1/\eps^2$ pole.
In the limit $v_{11}\rightarrow 0$, we have $\omega_{13}^+\rightarrow 1$ and $\omega_{13}^-\rightarrow 0$.
Analogously, in the limit $v_{22}\rightarrow 0$, it is $\omega_{24}^+\rightarrow 1$ and $\omega_{24}^-\rightarrow 0$.
Hence, a double massless pole term $\delta(v_{11})\,\delta(v_{22})/\eps^2$ will receive a contribution from the $\eps^2$ coefficient function in the double massless limit.
The specific value required for $f_1$ is $f_1(1,0)=-2\,\zeta_3$, where $\zeta_3$ denotes Ap\' ery`s constant ${\zeta_3=\sum_{n=1}^\infty 1/n^3}$.

If one is interested in only a single massless limit, say $v_{11}\rightarrow 0$ while $v_{22}$ stays finite, we may expand $v_{22}^{-\eps}$ allowing for some explicit simplifications of logarithms.
In this case, we find the representation
\begin{align}
&I^{(2)}_{1,1}(v_{12},v_{11},v_{22};\eps)=\frac{\pi}{\sqrt{X}}\left\lbrace-\frac{1}{\eps}-2\log\!\left(2\nu\right)
\right.
\nonumber\\
&\left.
\quad+\log\!\left(\frac{2 v_{12}(v_{12}+\sqrt{X})}{v_{22}}-v_{11}\right)
+2\eps\left(\dilog\!\left(\omega_{24}^+\right)
\right.\right.
\nonumber\\
&\left.\left.
\quad+\dilog\!\left(\omega_{24}^-\right)+\dilog\!\left(1-\frac{2v_{12}-v_{11}-v_{22}}{X}\right)
\right.
\right.
\\
&\left.\left.
\quad+\frac{1}{4}\log^2\!\left(\frac{v_{22}}{v_{24}^2}\right)\right)+\mathcal{O}(\eps^2)
+v_{11}^{-\eps}\left[\frac{1}{\eps}+2\log\left(2\nu\right)
\right.\right.
\nonumber\\
&\left.\left.
\quad+2\eps\left(\dilog\!\left(\omega_{13}^+\right)+\dilog\!\left(\omega_{13}^-\right)+\log^2\!\left(2\nu\right)\right)+\mathcal{O}(\eps^2)\vphantom{\frac{1}{\eps}}\right]
\right\rbrace,
\nonumber
\end{align}
with the abbreviation ${\nu\,=\,1\,-\,v_{11}/(2\,v_{14})}$, where ${v_{14}=\sqrt{X}(v_{12}-v_{11}+\sqrt{X})/(2 v_{12}-v_{11}-v_{22})}$.
The asymptotic form for $v_{22}\rightarrow 0$ at finite $v_{11}$ is the same upon interchanging $v_{11}\leftrightarrow v_{22}$.
\section{Conclusion}
\label{sec: Conclusion}
We have established $\eps$-expansions with manifest small-mass asymptotics for all massive angular master integrals with up to two denominators.
The main results of this paper are the asymptotic expansions of the
\begin{itemize}
\item massive integral $I^{(1)}_1$ in  Eq{.}\,\eqref{eq: Asymptotic form of one denominator integral},
\item single-massive integral $I^{(1)}_{1,1}$ in Eq{.}\,\eqref{eq: Asymptotic form of single massive integral},
\item double-massive integral $I^{(2)}_{1,1}$ in Eq{.}\,\eqref{eq: Asymptotic form of double massive integral}.
\end{itemize}

By means of recursion relations derived from IBP identities these results extend to all two-denominator angular integrals with integer coefficients.
In the construction of the asymptotic expansion the two-point splitting lemma proved to be an immensely useful tool.
It allowed for the extraction of the massless limits in terms of suitable massless angular integrals.

\appendix*
\section{Order \boldmath{$\eps^2$} coefficients of the double-massive integral}
\label{app: order eps^2 coefficients}
The explicit form of the order $\eps^2$ coefficient functions $f_0$ and $f_1$ of the massless respectively single-massive two-denominator integrals appearing in Eq{.}\,\eqref{eq: Asymptotic form of double massive integral} are
%\begin{align}
%f_0(v)=&-\trilog\!\left(1-\frac{v}{2}\right)-\trilog\!\left(\frac{v}{2}\right)+\dilog\!\left(\frac{v}{2}\right)\log\!\left(\frac{v}{2}\right)
%\nonumber
%\\
%&+\frac{1}{2}\log\!\left(1-\frac{v}{2}\right)\log^2\!\left(\frac{v}{2}\right)+\zeta_3
%\end{align}
\begin{align}
f_0(v)=\trilog\!\left(1-\frac{2}{v}\right)-\dilog\!\left(1-\frac{v}{2}\right)\log\frac{v}{2}-\frac{1}{6}\log^3\frac{v}{2}
\end{align}
and
\begin{align}
&f_1(\omega^+,\omega^-)=2\,\doublenielsentrilog(\omega^+,\omega^-)-2\,\trilog(\omega^+)
\nonumber\\
&
\qquad+2\,\trilog\!\left(\frac{\omega^+}{\omega^+-1}\right)
-2\,\dilog(\omega^+)\log(1-\omega^+)
\nonumber\\
&
\qquad-\frac{1}{3}\log^3(1-\omega^+)
+(\omega^+\leftrightarrow\omega^-)\,,
\end{align}
where $\doublenielsentrilog$ denotes the double Nielsen polylogarithm \citep{Lyubovitskij:2021}
\begin{align}
\doublenielsentrilog(x,y)=\int_0^1\frac{\dx t}{t}\log(1-x\,t)\log(1-y\,t)\,.
\label{eq: S111 integral}
\end{align}
This generalized polylogarithm of weight $3$ and depth $2$ can be expressed in terms of classical polylogarithms \cite{Frellesvig:2016,Duhr:2019}.
For $0<y<x<1$ it holds
\begin{align}
&\doublenielsentrilog(x,y)=
-\trilog\!\left(1-x\right)
-\trilog\!\left(1-y\right)
+\trilog\!\left(\frac{y}{x}\right)
\nonumber
\\
&\,\,+\trilog\!\left(\frac{1-x}{1-y}\right)
-\trilog\!\left(\frac{y(1-x)}{x(1-y)}\right)
-\dilog(y)\log(1-x)
\nonumber
\\
&\,\,
+\dilog(1-x)\log(1-y)
+\dilog\!\left(\frac{y}{x}\right)\log\!\left(\frac{1-x}{1-y}\right)
\\
&\,\,-\frac{1}{2}\log^2(1-y)\log\!\left(\frac{y}{x}\right)
+\frac{\pi^2}{6}\log(1-y)+\zeta_3\,.
\nonumber
\end{align}
The specific value needed for the double massless limit is $\doublenielsentrilog(1,0)=0$, which can be trivially read off from the integral representation \eqref{eq: S111 integral}.

\begin{acknowledgments}
The author thanks Mathias Butensch\"on for discussions stimulating this research. I am thankful to Juliane Haug, Timo Schreyer, and Werner Vogelsang for proofreading.
This work has been 
supported by Deutsche Forschungsgemeinschaft (DFG) through the Research Unit FOR 2926 (project 409651613).
\end{acknowledgments}

% Create the reference section using BibTeX:
\bibliography{Asymptotic_behavior_of_angular_integrals.bib}

\end{document}